\documentclass[reprint,amssymb, amsmath, aps, superscriptaddress, showpacs, footinbib, prl]{revtex4-1}
\usepackage{graphicx}
\usepackage{color}

\begin{document}

{\bf Comment on ``Unconventional Magnetism in a Nitrogen-Containing Analog of Cupric Oxide''} 
\smallskip

In a recent Letter, Zorko~\textit{et al.}~\cite{zorko} report on an ``unexpected inhomogeneous magnetism'' related to ``a peculiar fragility'' of the resonating-valence-bond (RVB) electronic state in the spin-$\frac12$ quantum magnet CuNCN. Here, we show that: i) the spin model of the anisotropic triangular lattice (ATL), as proposed in Ref.~\cite{zorko}, does not apply to CuNCN; ii) this model does not support the proclaimed RVB ground state in the relevant parameter range; iii) the magnetic state of CuNCN is not necessarily inhomogeneous. 

\begin{figure}[!b]
\includegraphics{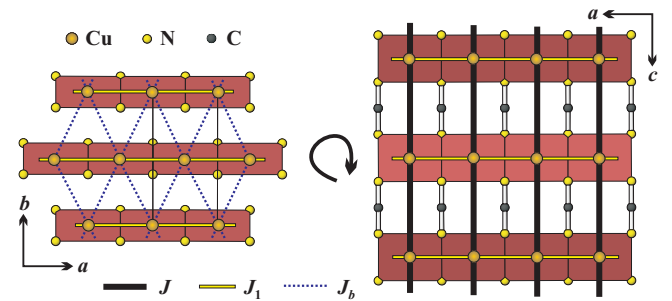}
\caption{\label{fig}
Microscopic magnetic model of CuNCN: anisotropic triangular lattice (ATL) proposed in Ref.~\cite{zorko} (left panel), and uniform spin chains running along the $c$ direction, according to band structure calculations~\cite{tsirlin2010} (right panel). The leading exchange coupling is $J\simeq 2500$~K compared to $J_1\simeq -500$~K and $J_b<2$~K.
}
\end{figure}

In Ref.~\cite{zorko}, the ATL geometry is derived from the two shortest Cu--Cu distances in the crystal structure (Fig.~\ref{fig}, left panel) and does not account for the $\pi$-conjugated NCN groups that mediate strong long-range superexchange interactions. The sizable dispersion of Cu $3d$ bands along the $c$ direction (Fig.~3 in Ref.~\cite{tsirlin2010}) indeed implies a large electron hopping and, consequently, a strong antiferromagnetic (AFM) exchange $J\simeq 2500$~K mediated by the NCN groups (Fig.~\ref{fig}, right panel). The similar NCN-driven long-range coupling has been reported in closely related transition-metal carbodiimides $M$NCN with $M$ = Fe, Co, and Ni~\cite{xiang2010,tsirlin2011}. The strong coupling $J$ is by far the leading energy scale of the system that does not simply modify the ATL physics, but essentially changes the microscopic scenario. The quasi-two-dimensional $J_1-J_b$ problem in the $ab$ plane is replaced by the quasi-one-dimensional (1D) lattice of spin chains running along the $c$ direction (see Fig.~\ref{fig}). 

Even if the leading coupling $J$ along the $c$ direction is ignored, the physics in the $ab$ plane does not support the proclaimed RVB scenario. According to the precise numerical study based on the coupled-cluster method (CCM)~\cite{bishop2009}, the phase space of the spin-$\frac12$ Heisenberg model on the ATL lattice features long-range-ordered (LRO) phases only. Other theoretical studies consistently restrict the \textit{possible} parameter range of the proposed gapped RVB-type phase to $1.2<J_1/J_b<1.6$~\cite{hauke2011}, while Zorko~\textit{et~al.}~\cite{zorko} estimate $J_1=2300$~K, $J_b=540$~K, and $J_1/J_b\simeq 4.3$, which is well inside the collinear stripe-ordered phase of the ATL model~\cite{bishop2009,starykh2007}. Moreover, the leading coupling $J_1$ of 2300~K is found for the Cu--N--Cu angle of $96.75^{\circ}$, where a ferromagnetic or weakly AFM interaction is expected from Goodenough-Kanamori rules. An RVB scenario also contradicts the computational estimates of Ref.~\cite{tsirlin2010} ($J_1\simeq -500$~K, $J_b\simeq 10$~K). 

The RVB ground state lacks static magnetic fields and shows short-range spin correlations only. However, both muon spin rotation ($\mu$SR) and nuclear magnetic resonance (NMR) experiments evidence the formation of static fields below $70-80$~K. To circumvent this discrepancy, the authors of Ref.~\cite{zorko} suggest the ``peculiar fragility'' of the RVB state, which is allegedly destroyed by the muons and by the magnetic field of 9.4~T in the $\mu$SR and NMR experiments, respectively. In this situation, only the electron spin resonance (ESR) measurement performed at a lower field of 0.1~T could access the true ground state, but the ESR response is at best ambiguous and does not show any specific footprints of the RVB scenario. The decrease in the ESR intensity below 70~K can be evenly understood as the onset of the conventional long-range magnetic order that is indeed expected from microscopic considerations~\cite{tsirlin2010}.

Starting from this microscopic viewpoint, we interpret the static magnetic fields observed in the $\mu$SR and NMR experiments as further signatures of the LRO state below 70~K. In $\mu$SR, the broad distribution of internal fields can be ascribed to several inequivalent muon stop sites, while the low value of the stretch exponent (inset to the right panel of Fig.~3 in Ref.~\cite{zorko}) is well understood in terms of two distinct relaxation times~\cite{johnston}. Therefore, neither $\mu$SR nor NMR show unambiguous signatures of an inhomogeneous magnetic state. The experimental data can be thus reconciled in terms of the conventional LRO scenario and do not require an unnecessary assumption of an unconventional fragile electronic state.

In conclusion, the RVB ground state of CuNCN is not supported by direct experimental evidence, and contradicts both: i) well-established numerical results for the ATL model~\cite{bishop2009,starykh2007,hauke2011}; ii) microscopic description of the compound in terms of its electronic structure~\cite{tsirlin2010}. Further work on CuNCN is presently underway to better understand experimental observations.

\smallskip
\indent
{Alexander A. Tsirlin and Helge Rosner}\\ 
\indent 
{MPI-CPfS, Dresden, Germany}

{\bf Addendum to the Comment} 
\smallskip

The magnetic state of CuNCN below 70~K has been a matter of debate in the recent literature~\cite{tsirlin2010-2,liu2008,xiang2009}. Here, we provide extra comments on publications preceding the Letter by Zorko~\textit{et al.}~\cite{zorko2}, and discuss their results in the context of the microscopic magnetic model from Ref.~\cite{tsirlin2010-2}. We focus on two main problems: i) the microscopic magnetic model of CuNCN; ii) experimental observation of the potential long-range-ordered (LRO) magnetic state in this compound. The following comments extend our published work~\cite{tsirlin2010-2}, and clarify its relation to experimental results presented in Refs.~\cite{liu2008,xiang2009}.

\begin{figure}
\includegraphics{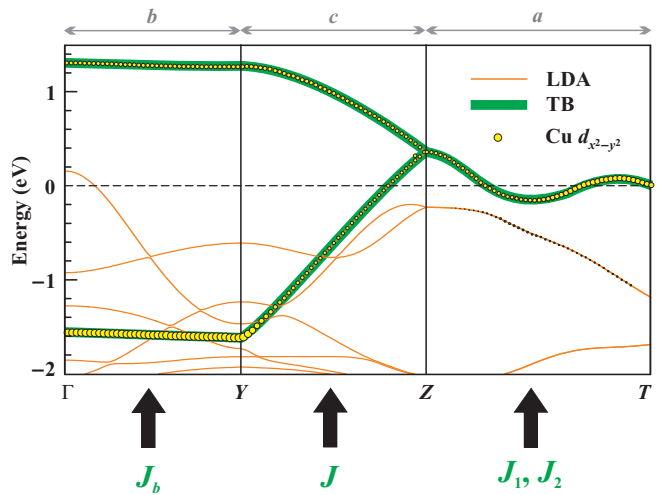}
\caption{\label{fig:band}
Left panel: band structure of CuNCN calculated in local density approximation (LDA) and the fit with the tight-binding (TB) model shown in thin orange and thick green lines, respectively. Circles denote the contribution of the half-filled Cu $d_{x^2-y^2}$ orbital. The large dispersion along $Y-Z$ identifies the leading AFM coupling $J$ along $c$, while flat bands along $\Gamma-Y$ clearly show the very weak AFM coupling $J_b$. Finally, the complex dispersion along $Z-T$ suggest the competition of the nearest-neighbor ($J_1$) and next-nearest-neighbor ($J_2$) couplings along the structural CuN$_2$ chains ($a$ direction). Note that in this analysis we start from the uncorrelated (metallic) LDA band structure and obtain the insulating solution after mapping the LDA band structure onto an effective Hubbard model.
}
\end{figure}

Reliable information on the electronic structure is a necessary prerequisite of the microscopic magnetic model. The electronic structure of CuNCN has been first considered by Liu~\textit{et al.}~\cite{liu2008}, who performed density functional theory (DFT)+$U$ calculations and arrived at the anisotropic triangular lattice (ATL) spin model (see Fig.~1 of the Comment). Their results apparently contradict band dispersions shown in Fig.~\ref{fig:band}, because the strong antiferromagnetic (AFM) coupling $J_b$ is inconsistent with the nearly flat Cu bands along the $b^*$ direction of the reciprocal space. By contrast, the large band dispersion along $c$ should lead to a strong AFM coupling, which is, however, not taken into account in the ATL model of Ref.~\cite{liu2008}. The ATL model was indeed disproved, but -- surprisingly -- for a totally different reason, when Xiang~\textit{et al.}~\cite{xiang2009} claimed a non-magnetic state of Cu$^{+2}$ based on new band structure calculations. 

The rather unconventional conclusion on the non-magnetic Cu$^{+2}$ in an insulating compound contradicts the ESR response observed by Zorko~\textit{et al.}~\cite{zorko2}. This Letter further refers to the ATL model of Ref.~\cite{liu2008}, although its results were disproved in Ref.~\cite{xiang2009} by the same group. To the best of our knowledge, the relationship between the mutually contradicting results of Refs.~\cite{liu2008,xiang2009} and the model assumptions of Ref.~\cite{zorko2} has not been clarified.

Once all relevant interactions are taken into account, DFT-based calculations provide a consistent microscopic picture that is robust with respect to subtle technical details, such as the treatment of strong electronic correlations on the Cu$^{+2}$ site. The conclusion on the strong AFM coupling $J$ mediated by the NCN groups stems from the large band dispersion along the $Y-Z$ direction of the reciprocal space (see Fig.~\ref{fig:band}). By contrast, the dispersion along $\Gamma-Y$ is vanishingly small, thus suggesting the negligible role of $J_b$ that, according to Refs.~\cite{liu2008,zorko2}, is an integral part of the anisotropic triangular spin lattice. These results are robust with respect to a specific choice of the exchange-correlation potential, and have been further confirmed by DFT+$U$ calculations~\cite{tsirlin2010-2}. The microscopic analysis also puts forward the close similarity between CuNCN and Cu$^{+2}$ oxides, because the relevant ligand states of the NCN unit are formed by $2p$ orbitals of nitrogen, similar to oxygen $2p$ orbitals in cuprates. Considering the long-standing experience of successful applications of DFT and DFT+$U$ to Cu$^{+2}$ oxides~\cite{[{For example: }][{}]valenti2002,*johannes2006,*mazurenko2008,*mentre2009,*cu2v2o7,*dioptase}, the microscopic scenario for CuNCN is reliable. In a carbodiimide system, an unexpected failure of DFT-based methods that have proved their efficiency for oxides can be ruled out. Therefore, the leading AFM coupling $J$ has to be taken into consideration, and the ATL spin model of CuNCN should be replaced by the microscopic-based one (Fig.~1 of the Comment).

Concerning experimental results, the onset of static magnetic fields at 70~K is neither accompanied by an anomaly in the specific heat, nor observed by neutron scattering. Zorko~\textit{et al.}~\cite{zorko2} consider these experimental facts as additional arguments favoring the ``unexpected inhomogeneous magnetic state'' rather than the conventional LRO in CuNCN. The onset of the magnetic order without the corresponding specific-heat anomaly and neutron signal is surprising indeed, although not uncommon in low-dimensional spin-$\frac12$ quantum magnets with low ordered magnetic moments, caused by strong quantum fluctuations, and diminutively small magnetic entropy at the N\'eel temperature $T_N$. 

\begin{figure}
\includegraphics{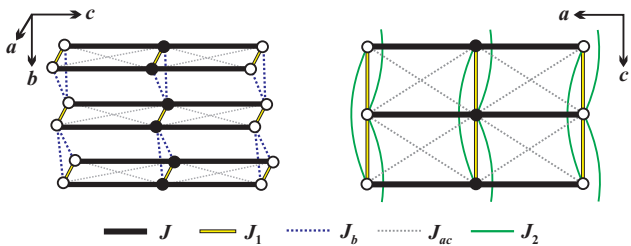}
\caption{\label{fig:lattice}
Spin lattice of CuNCN according to the computational results of Ref.~\cite{tsirlin2010-2}: $J\simeq 2500$~K, $J_1\simeq -500$~K, $J_2\simeq 100$~K, $J_{ac}\simeq 70$~K, and $J_b<2$~K. Open and filled circles denote up and down spins in the ground state spin configuration derived from CCM calculations. The ordered magnetic moment is about 0.4~$\mu_B$.
}
\end{figure}
\begin{figure}
\includegraphics{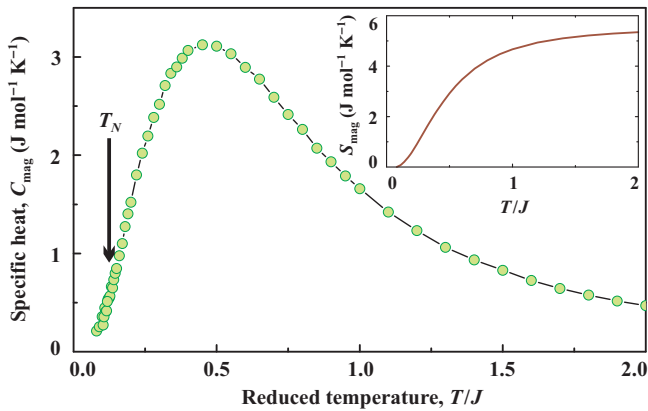}
\caption{\label{fig:heat}
Magnetic specific heat of CuNCN calculated with quantum Monte-Carlo simulations, and the respective magnetic entropy shown as an inset. Note the absence of any visible anomaly at the magnetic ordering transition $T_N/J\simeq 0.12$ due to the very low entropy available at $T_N$.
}
\end{figure}

To better understand the experimental results, we performed further calculations for the microscopic magnetic model of CuNCN established in Ref.~\cite{tsirlin2010-2}. Our quantum Monte-Carlo simulations~\cite{Note1} reveal $T_N/J\simeq 0.12$ and confirm the absence of any visible anomaly in the magnetic specific heat (Fig.~\ref{fig:heat}). Experimentally, the magnetic contribution to the specific heat is superimposed onto the lattice contribution (about 20 J~mol$^{-1}$~K$^{-1}$ at $T_N=70$~K) that further impedes the experimental observation of the magnetic transition. The small transition anomaly originates from the low magnetic entropy (below 3~\% of the full magnetic entropy $R\ln 2$) available at $T_N$. Similar results for the missing specific-heat anomaly at $T_N$ have been recently obtained for a family of quasi-two-dimensional spin-$\frac12$ quantum magnets~\cite{lancaster2007,*sengupta2003}.

The absence of the magnetic neutron scattering can be ascribed to a spin-liquid ground state, which is devoid of the long-range magnetic order, or to the LRO state with the small magnetic moment that lies below the sensitivity threshold of the neutron-scattering experiment. Applying the coupled-cluster method to our DFT-based microscopic model, we establish the stripe AFM ordering pattern (Fig.~\ref{fig:lattice}) and estimate the ordered magnetic moment of about 0.4~$\mu_B$ in CuNCN. An experimental observation of such a low magnetic moment remains challenging and requires instruments with excellent sensitivity and, preferably, single crystals, while in Refs.~\cite{liu2008} and~\cite{xiang2009} none of these conditions were fulfilled. We believe that the interpretation of the neutron data is premature, because neither Liu~\textit{et al.}~\cite{liu2008} nor Xiang~\textit{et al.}~\cite{xiang2009} compared the sensitivity of their neutron measurements to the magnitude of the anticipated magnetic scattering in CuNCN. The systems showing the LRO magnetic state and lacking any magnetic scattering observable in a powder neutron experiment include, for instance, Sr$_2$CuO$_3$~\cite{kojima1997,*ami1995} and PbVO$_3$~\cite{shpanchenko2004,*tsirlin2008,*oka2008}. In quantum magnets, this situation is not uncommon, and thus the reported neutron experiments alone can not be a decisive test for the formation of the LRO magnetic state.

We would like to acknowledge Ronald Zinke and Johannes Richter for providing the CCM results.

\smallskip
\indent
{Alexander A. Tsirlin and Helge Rosner}\\ 
\indent 
{MPI-CPfS, Dresden, Germany}


\vspace{-0.5cm}
\begin{thebibliography}{7}
\vspace{-0.45cm}
\bibitem{zorko} A. Zorko~\textit{et al.}, Phys. Rev. Lett. \textbf{107}, 047208 (2011).
\bibitem{tsirlin2010} A. Tsirlin and H. Rosner, Phys. Rev. B \textbf{81}, 024424 (2010).
\bibitem{xiang2010} H. Xiang~\textit{et al.}, J. Phys. Chem. A \textbf{114}, 12345 (2010).
\bibitem{tsirlin2011} A. A. Tsirlin and H. Rosner, arXiv:1106.3665.
\bibitem{bishop2009} R. F. Bishop~\textit{et al.}, Phys. Rev. B \textbf{79}, 174405 (2009).
\bibitem{starykh2007} O. Starykh~\textit{et al.}, Phys. Rev. Lett. \textbf{98}, 077205 (2007).
\bibitem{hauke2011} P. Hauke~\textit{et al.}, New J. Phys. \textbf{13}, 075017 (2011).
\bibitem{johnston} D. C. Johnston, Phys. Rev. B \textbf{74}, 184430 (2006).
\end{thebibliography}

\begin{thebibliography}{18}%
\makeatletter
\providecommand \@ifxundefined [1]{%
 \@ifx{#1\undefined}
}%
\providecommand \@ifnum [1]{%
 \ifnum #1\expandafter \@firstoftwo
 \else \expandafter \@secondoftwo
 \fi
}%
\providecommand \@ifx [1]{%
 \ifx #1\expandafter \@firstoftwo
 \else \expandafter \@secondoftwo
 \fi
}%
\providecommand \natexlab [1]{#1}%
\providecommand \enquote  [1]{``#1''}%
\providecommand \bibnamefont  [1]{#1}%
\providecommand \bibfnamefont [1]{#1}%
\providecommand \citenamefont [1]{#1}%
\providecommand \href@noop [0]{\@secondoftwo}%
\providecommand \href [0]{\begingroup \@sanitize@url \@href}%
\providecommand \@href[1]{\@@startlink{#1}\@@href}%
\providecommand \@@href[1]{\endgroup#1\@@endlink}%
\providecommand \@sanitize@url [0]{\catcode `\\12\catcode `\$12\catcode
  `\&12\catcode `\#12\catcode `\^12\catcode `\_12\catcode `\%12\relax}%
\providecommand \@@startlink[1]{}%
\providecommand \@@endlink[0]{}%
\providecommand \url  [0]{\begingroup\@sanitize@url \@url }%
\providecommand \@url [1]{\endgroup\@href {#1}{\urlprefix }}%
\providecommand \urlprefix  [0]{URL }%
\providecommand \Eprint [0]{\href }%
\providecommand \doibase [0]{http://dx.doi.org/}%
\providecommand \selectlanguage [0]{\@gobble}%
\providecommand \bibinfo  [0]{\@secondoftwo}%
\providecommand \bibfield  [0]{\@secondoftwo}%
\providecommand \translation [1]{[#1]}%
\providecommand \BibitemOpen [0]{}%
\providecommand \bibitemStop [0]{}%
\providecommand \bibitemNoStop [0]{.\EOS\space}%
\providecommand \EOS [0]{\spacefactor3000\relax}%
\providecommand \BibitemShut  [1]{\csname bibitem#1\endcsname}%
\let\auto@bib@innerbib\@empty
\bibitem [{\citenamefont {Tsirlin}\ and\ \citenamefont
  {Rosner}(2010)}]{tsirlin2010-2}%
  \BibitemOpen
  \bibfield  {author} {\bibinfo {author} {\bibfnamefont {A.~A.}\ \bibnamefont
  {Tsirlin}}\ and\ \bibinfo {author} {\bibfnamefont {H.}~\bibnamefont
  {Rosner}},\ }\href@noop {} {\bibfield  {journal} {\bibinfo  {journal} {Phys.
  Rev. B}\ }\textbf {\bibinfo {volume} {81}},\ \bibinfo {pages} {024424}
  (\bibinfo {year} {2010})}\BibitemShut {NoStop}%
\bibitem [{\citenamefont {Liu}\ \emph {et~al.}(2008)\citenamefont {Liu},
  \citenamefont {Dronskowski}, \citenamefont {Kremer}, \citenamefont {Ahrens},
  \citenamefont {Lee},\ and\ \citenamefont {Whangbo}}]{liu2008}%
  \BibitemOpen
  \bibfield  {author} {\bibinfo {author} {\bibfnamefont {X.}~\bibnamefont
  {Liu}}, \bibinfo {author} {\bibfnamefont {R.}~\bibnamefont {Dronskowski}},
  \bibinfo {author} {\bibfnamefont {R.~K.}\ \bibnamefont {Kremer}}, \bibinfo
  {author} {\bibfnamefont {M.}~\bibnamefont {Ahrens}}, \bibinfo {author}
  {\bibfnamefont {C.}~\bibnamefont {Lee}}, \ and\ \bibinfo {author}
  {\bibfnamefont {M.-H.}\ \bibnamefont {Whangbo}},\ }\href@noop {} {\bibfield
  {journal} {\bibinfo  {journal} {J. Phys. Chem. C}\ }\textbf {\bibinfo
  {volume} {112}},\ \bibinfo {pages} {11013} (\bibinfo {year}
  {2008})}\BibitemShut {NoStop}%
\bibitem [{\citenamefont {Xiang}\ \emph {et~al.}(2009)\citenamefont {Xiang},
  \citenamefont {Liu},\ and\ \citenamefont {Dronskowski}}]{xiang2009}%
  \BibitemOpen
  \bibfield  {author} {\bibinfo {author} {\bibfnamefont {H.}~\bibnamefont
  {Xiang}}, \bibinfo {author} {\bibfnamefont {X.}~\bibnamefont {Liu}}, \ and\
  \bibinfo {author} {\bibfnamefont {R.}~\bibnamefont {Dronskowski}},\
  }\href@noop {} {\bibfield  {journal} {\bibinfo  {journal} {J. Phys. Chem. C}\
  }\textbf {\bibinfo {volume} {113}},\ \bibinfo {pages} {18891} (\bibinfo
  {year} {2009})}\BibitemShut {NoStop}%
\bibitem [{\citenamefont {Zorko}\ \emph {et~al.}(2011)\citenamefont {Zorko},
  \citenamefont {Jegli{\v c}}, \citenamefont {Poto{\v c}nik}, \citenamefont
  {Ar{\v c}on}, \citenamefont {Bal{\v c}ytis}, \citenamefont {Jagli{\v
  c}i{\'c}}, \citenamefont {Liu}, \citenamefont {Tchougr\'eeff},\ and\
  \citenamefont {Dronskowski}}]{zorko2}%
  \BibitemOpen
  \bibfield  {author} {\bibinfo {author} {\bibfnamefont {A.}~\bibnamefont
  {Zorko}}, \bibinfo {author} {\bibfnamefont {P.}~\bibnamefont {Jegli{\v c}}},
  \bibinfo {author} {\bibfnamefont {A.}~\bibnamefont {Poto{\v c}nik}}, \bibinfo
  {author} {\bibfnamefont {D.}~\bibnamefont {Ar{\v c}on}}, \bibinfo {author}
  {\bibfnamefont {A.}~\bibnamefont {Bal{\v c}ytis}}, \bibinfo {author}
  {\bibfnamefont {Z.}~\bibnamefont {Jagli{\v c}i{\'c}}}, \bibinfo {author}
  {\bibfnamefont {X.}~\bibnamefont {Liu}}, \bibinfo {author} {\bibfnamefont
  {A.~L.}\ \bibnamefont {Tchougr\'eeff}}, \ and\ \bibinfo {author}
  {\bibfnamefont {R.}~\bibnamefont {Dronskowski}},\ }\href@noop {} {\bibfield
  {journal} {\bibinfo  {journal} {Phys. Rev. Lett.}\ }\textbf {\bibinfo
  {volume} {107}},\ \bibinfo {pages} {047208} (\bibinfo {year}
  {2011})}\BibitemShut {NoStop}%
\bibitem [{\citenamefont {Valenti}\ \emph {et~al.}(2002)\citenamefont
  {Valenti}, \citenamefont {Saha-Dasgupta},\ and\ \citenamefont
  {Gros}}]{valenti2002}%
  \BibitemOpen
  \bibfield  {author} {\bibinfo {author} {\bibfnamefont {R.}~\bibnamefont
  {Valenti}}, \bibinfo {author} {\bibfnamefont {T.}~\bibnamefont
  {Saha-Dasgupta}}, \ and\ \bibinfo {author} {\bibfnamefont {C.}~\bibnamefont
  {Gros}},\ }\href@noop {} {\bibfield  {journal} {\bibinfo  {journal} {Phys.
  Rev. B}\ }\textbf {\bibinfo {volume} {66}},\ \bibinfo {pages} {054426}
  (\bibinfo {year} {2002})}\BibitemShut {NoStop}%
\bibitem [{\citenamefont {Johannes}\ \emph {et~al.}(2006)\citenamefont
  {Johannes}, \citenamefont {Richter}, \citenamefont {Drechsler},\ and\
  \citenamefont {Rosner}}]{johannes2006}%
  \BibitemOpen
  \bibfield  {author} {\bibinfo {author} {\bibfnamefont {M.~D.}\ \bibnamefont
  {Johannes}}, \bibinfo {author} {\bibfnamefont {J.}~\bibnamefont {Richter}},
  \bibinfo {author} {\bibfnamefont {S.-L.}\ \bibnamefont {Drechsler}}, \ and\
  \bibinfo {author} {\bibfnamefont {H.}~\bibnamefont {Rosner}},\ }\href@noop {}
  {\bibfield  {journal} {\bibinfo  {journal} {Phys. Rev. B}\ }\textbf {\bibinfo
  {volume} {74}},\ \bibinfo {pages} {174435} (\bibinfo {year}
  {2006})}\BibitemShut {NoStop}%
\bibitem [{\citenamefont {Mazurenko}\ \emph {et~al.}(2008)\citenamefont
  {Mazurenko}, \citenamefont {Skornyakov}, \citenamefont {Anisimov},\ and\
  \citenamefont {Mila}}]{mazurenko2008}%
  \BibitemOpen
  \bibfield  {author} {\bibinfo {author} {\bibfnamefont {V.~V.}\ \bibnamefont
  {Mazurenko}}, \bibinfo {author} {\bibfnamefont {S.~L.}\ \bibnamefont
  {Skornyakov}}, \bibinfo {author} {\bibfnamefont {V.~I.}\ \bibnamefont
  {Anisimov}}, \ and\ \bibinfo {author} {\bibfnamefont {F.}~\bibnamefont
  {Mila}},\ }\href@noop {} {\bibfield  {journal} {\bibinfo  {journal} {Phys.
  Rev. B}\ }\textbf {\bibinfo {volume} {78}},\ \bibinfo {pages} {195110}
  (\bibinfo {year} {2008})}\BibitemShut {NoStop}%
\bibitem [{\citenamefont {Mentr\'e}\ \emph {et~al.}(2009)\citenamefont
  {Mentr\'e}, \citenamefont {Janod}, \citenamefont {Rabu}, \citenamefont
  {Hennion}, \citenamefont {Leclercq-Hugeux}, \citenamefont {Kang},
  \citenamefont {Lee}, \citenamefont {Whangbo},\ and\ \citenamefont
  {Petit}}]{mentre2009}%
  \BibitemOpen
  \bibfield  {author} {\bibinfo {author} {\bibfnamefont {O.}~\bibnamefont
  {Mentr\'e}}, \bibinfo {author} {\bibfnamefont {E.}~\bibnamefont {Janod}},
  \bibinfo {author} {\bibfnamefont {P.}~\bibnamefont {Rabu}}, \bibinfo {author}
  {\bibfnamefont {M.}~\bibnamefont {Hennion}}, \bibinfo {author} {\bibfnamefont
  {F.}~\bibnamefont {Leclercq-Hugeux}}, \bibinfo {author} {\bibfnamefont
  {J.}~\bibnamefont {Kang}}, \bibinfo {author} {\bibfnamefont {C.}~\bibnamefont
  {Lee}}, \bibinfo {author} {\bibfnamefont {M.-H.}\ \bibnamefont {Whangbo}}, \
  and\ \bibinfo {author} {\bibfnamefont {S.}~\bibnamefont {Petit}},\
  }\href@noop {} {\bibfield  {journal} {\bibinfo  {journal} {Phys. Rev. B}\
  }\textbf {\bibinfo {volume} {80}},\ \bibinfo {pages} {180413(R)} (\bibinfo
  {year} {2009})}\BibitemShut {NoStop}%
\bibitem [{\citenamefont {Tsirlin}\ \emph {et~al.}(2010)\citenamefont
  {Tsirlin}, \citenamefont {Janson},\ and\ \citenamefont {Rosner}}]{cu2v2o7}%
  \BibitemOpen
  \bibfield  {author} {\bibinfo {author} {\bibfnamefont {A.~A.}\ \bibnamefont
  {Tsirlin}}, \bibinfo {author} {\bibfnamefont {O.}~\bibnamefont {Janson}}, \
  and\ \bibinfo {author} {\bibfnamefont {H.}~\bibnamefont {Rosner}},\
  }\href@noop {} {\bibfield  {journal} {\bibinfo  {journal} {Phys. Rev. B}\
  }\textbf {\bibinfo {volume} {82}},\ \bibinfo {pages} {144416} (\bibinfo
  {year} {2010})}\BibitemShut {NoStop}%
\bibitem [{\citenamefont {Janson}\ \emph {et~al.}(2010)\citenamefont {Janson},
  \citenamefont {Tsirlin}, \citenamefont {Schmitt},\ and\ \citenamefont
  {Rosner}}]{dioptase}%
  \BibitemOpen
  \bibfield  {author} {\bibinfo {author} {\bibfnamefont {O.}~\bibnamefont
  {Janson}}, \bibinfo {author} {\bibfnamefont {A.~A.}\ \bibnamefont {Tsirlin}},
  \bibinfo {author} {\bibfnamefont {M.}~\bibnamefont {Schmitt}}, \ and\
  \bibinfo {author} {\bibfnamefont {H.}~\bibnamefont {Rosner}},\ }\href@noop {}
  {\bibfield  {journal} {\bibinfo  {journal} {Phys. Rev. B}\ }\textbf {\bibinfo
  {volume} {82}},\ \bibinfo {pages} {014424} (\bibinfo {year}
  {2010})}\BibitemShut {NoStop}%
\bibitem [{Note1()}]{Note1}%
  \BibitemOpen
  \bibinfo {note} {To avoid sign problem for a frustrated spin lattice, we
  consider the uniform coupling along the $b$ direction instead of the
  frustrated coupling $J_b$, and omit the second-neighbor coupling $J_2$ along
  $a$. These modifications should reduce quantum fluctuations and,
  consequently, enhance the transition temperature as well as the specific-heat
  anomaly. Therefore, our Monte-Carlo simulations provide the upper estimate
  for the $T_N$ and transition entropy in CuNCN.}\BibitemShut {Stop}%
\bibitem [{\citenamefont {Lancaster}\ \emph {et~al.}(2007)\citenamefont
  {Lancaster}, \citenamefont {Blundell}, \citenamefont {Brooks}, \citenamefont
  {Baker}, \citenamefont {Pratt}, \citenamefont {Manson}, \citenamefont
  {Conner}, \citenamefont {Xiao}, \citenamefont {Landee}, \citenamefont
  {Chaves}, \citenamefont {Soriano}, \citenamefont {Novak}, \citenamefont
  {Papageorgiou}, \citenamefont {Bianchi}, \citenamefont {Herrmannsd\"orfer},
  \citenamefont {Wosnitza},\ and\ \citenamefont {Schlueter}}]{lancaster2007}%
  \BibitemOpen
  \bibfield  {author} {\bibinfo {author} {\bibfnamefont {T.}~\bibnamefont
  {Lancaster}}, \bibinfo {author} {\bibfnamefont {S.~J.}\ \bibnamefont
  {Blundell}}, \bibinfo {author} {\bibfnamefont {M.~L.}\ \bibnamefont
  {Brooks}}, \bibinfo {author} {\bibfnamefont {P.~J.}\ \bibnamefont {Baker}},
  \bibinfo {author} {\bibfnamefont {F.~L.}\ \bibnamefont {Pratt}}, \bibinfo
  {author} {\bibfnamefont {J.~L.}\ \bibnamefont {Manson}}, \bibinfo {author}
  {\bibfnamefont {M.~M.}\ \bibnamefont {Conner}}, \bibinfo {author}
  {\bibfnamefont {F.}~\bibnamefont {Xiao}}, \bibinfo {author} {\bibfnamefont
  {C.~P.}\ \bibnamefont {Landee}}, \bibinfo {author} {\bibfnamefont {F.~A.}\
  \bibnamefont {Chaves}}, \bibinfo {author} {\bibfnamefont {S.}~\bibnamefont
  {Soriano}}, \bibinfo {author} {\bibfnamefont {M.~A.}\ \bibnamefont {Novak}},
  \bibinfo {author} {\bibfnamefont {T.~P.}\ \bibnamefont {Papageorgiou}},
  \bibinfo {author} {\bibfnamefont {A.~D.}\ \bibnamefont {Bianchi}}, \bibinfo
  {author} {\bibfnamefont {T.}~\bibnamefont {Herrmannsd\"orfer}}, \bibinfo
  {author} {\bibfnamefont {J.}~\bibnamefont {Wosnitza}}, \ and\ \bibinfo
  {author} {\bibfnamefont {J.~A.}\ \bibnamefont {Schlueter}},\ }\href@noop {}
  {\bibfield  {journal} {\bibinfo  {journal} {Phys. Rev. B}\ }\textbf {\bibinfo
  {volume} {75}},\ \bibinfo {pages} {094421} (\bibinfo {year}
  {2007})}\BibitemShut {NoStop}%
\bibitem [{\citenamefont {Sengupta}\ \emph {et~al.}(2003)\citenamefont
  {Sengupta}, \citenamefont {Sandvik},\ and\ \citenamefont
  {Singh}}]{sengupta2003}%
  \BibitemOpen
  \bibfield  {author} {\bibinfo {author} {\bibfnamefont {P.}~\bibnamefont
  {Sengupta}}, \bibinfo {author} {\bibfnamefont {A.~W.}\ \bibnamefont
  {Sandvik}}, \ and\ \bibinfo {author} {\bibfnamefont {R.~R.~P.}\ \bibnamefont
  {Singh}},\ }\href@noop {} {\bibfield  {journal} {\bibinfo  {journal} {Phys.
  Rev. B}\ }\textbf {\bibinfo {volume} {68}},\ \bibinfo {pages} {094423}
  (\bibinfo {year} {2003})}\BibitemShut {NoStop}%
\bibitem [{\citenamefont {Kojima}\ \emph {et~al.}(1997)\citenamefont {Kojima},
  \citenamefont {Fudamoto}, \citenamefont {Larkin}, \citenamefont {Luke},
  \citenamefont {Merrin}, \citenamefont {Nachumi}, \citenamefont {Uemura},
  \citenamefont {Motoyama}, \citenamefont {Eisaki}, \citenamefont {Uchida},
  \citenamefont {Yamada}, \citenamefont {Endoh}, \citenamefont {Hosoya},
  \citenamefont {Sternlieb},\ and\ \citenamefont {Shirane}}]{kojima1997}%
  \BibitemOpen
  \bibfield  {author} {\bibinfo {author} {\bibfnamefont {K.~M.}\ \bibnamefont
  {Kojima}}, \bibinfo {author} {\bibfnamefont {Y.}~\bibnamefont {Fudamoto}},
  \bibinfo {author} {\bibfnamefont {M.}~\bibnamefont {Larkin}}, \bibinfo
  {author} {\bibfnamefont {G.~M.}\ \bibnamefont {Luke}}, \bibinfo {author}
  {\bibfnamefont {J.}~\bibnamefont {Merrin}}, \bibinfo {author} {\bibfnamefont
  {B.}~\bibnamefont {Nachumi}}, \bibinfo {author} {\bibfnamefont {Y.~J.}\
  \bibnamefont {Uemura}}, \bibinfo {author} {\bibfnamefont {N.}~\bibnamefont
  {Motoyama}}, \bibinfo {author} {\bibfnamefont {H.}~\bibnamefont {Eisaki}},
  \bibinfo {author} {\bibfnamefont {S.}~\bibnamefont {Uchida}}, \bibinfo
  {author} {\bibfnamefont {K.}~\bibnamefont {Yamada}}, \bibinfo {author}
  {\bibfnamefont {Y.}~\bibnamefont {Endoh}}, \bibinfo {author} {\bibfnamefont
  {S.}~\bibnamefont {Hosoya}}, \bibinfo {author} {\bibfnamefont {B.~J.}\
  \bibnamefont {Sternlieb}}, \ and\ \bibinfo {author} {\bibfnamefont
  {G.}~\bibnamefont {Shirane}},\ }\href@noop {} {\bibfield  {journal} {\bibinfo
   {journal} {Phys. Rev. Lett.}\ }\textbf {\bibinfo {volume} {78}},\ \bibinfo
  {pages} {1787} (\bibinfo {year} {1997})}\BibitemShut {NoStop}%
\bibitem [{\citenamefont {Ami}\ \emph {et~al.}(1995)\citenamefont {Ami},
  \citenamefont {Crawford}, \citenamefont {Harlow}, \citenamefont {Wang},
  \citenamefont {Johnston}, \citenamefont {Huang},\ and\ \citenamefont
  {Erwin}}]{ami1995}%
  \BibitemOpen
  \bibfield  {author} {\bibinfo {author} {\bibfnamefont {T.}~\bibnamefont
  {Ami}}, \bibinfo {author} {\bibfnamefont {M.~K.}\ \bibnamefont {Crawford}},
  \bibinfo {author} {\bibfnamefont {R.~L.}\ \bibnamefont {Harlow}}, \bibinfo
  {author} {\bibfnamefont {Z.~R.}\ \bibnamefont {Wang}}, \bibinfo {author}
  {\bibfnamefont {D.~C.}\ \bibnamefont {Johnston}}, \bibinfo {author}
  {\bibfnamefont {Q.}~\bibnamefont {Huang}}, \ and\ \bibinfo {author}
  {\bibfnamefont {R.~W.}\ \bibnamefont {Erwin}},\ }\href@noop {} {\bibfield
  {journal} {\bibinfo  {journal} {Phys. Rev. B}\ }\textbf {\bibinfo {volume}
  {51}},\ \bibinfo {pages} {5994} (\bibinfo {year} {1995})}\BibitemShut
  {NoStop}%
\bibitem [{\citenamefont {Shpanchenko}\ \emph {et~al.}(2004)\citenamefont
  {Shpanchenko}, \citenamefont {Chernaya}, \citenamefont {Tsirlin},
  \citenamefont {Chizhov}, \citenamefont {Sklovsky}, \citenamefont {Antipov},
  \citenamefont {Khlybov}, \citenamefont {Pomjakushin}, \citenamefont
  {Balagurov}, \citenamefont {Medvedeva}, \citenamefont {Kaul},\ and\
  \citenamefont {Geibel}}]{shpanchenko2004}%
  \BibitemOpen
  \bibfield  {author} {\bibinfo {author} {\bibfnamefont {R.~V.}\ \bibnamefont
  {Shpanchenko}}, \bibinfo {author} {\bibfnamefont {V.~V.}\ \bibnamefont
  {Chernaya}}, \bibinfo {author} {\bibfnamefont {A.~A.}\ \bibnamefont
  {Tsirlin}}, \bibinfo {author} {\bibfnamefont {P.~S.}\ \bibnamefont
  {Chizhov}}, \bibinfo {author} {\bibfnamefont {D.~E.}\ \bibnamefont
  {Sklovsky}}, \bibinfo {author} {\bibfnamefont {E.~V.}\ \bibnamefont
  {Antipov}}, \bibinfo {author} {\bibfnamefont {E.~P.}\ \bibnamefont
  {Khlybov}}, \bibinfo {author} {\bibfnamefont {V.}~\bibnamefont
  {Pomjakushin}}, \bibinfo {author} {\bibfnamefont {A.~M.}\ \bibnamefont
  {Balagurov}}, \bibinfo {author} {\bibfnamefont {J.~E.}\ \bibnamefont
  {Medvedeva}}, \bibinfo {author} {\bibfnamefont {E.~E.}\ \bibnamefont {Kaul}},
  \ and\ \bibinfo {author} {\bibfnamefont {C.}~\bibnamefont {Geibel}},\
  }\href@noop {} {\bibfield  {journal} {\bibinfo  {journal} {Chem. Mater.}\
  }\textbf {\bibinfo {volume} {16}},\ \bibinfo {pages} {3267} (\bibinfo {year}
  {2004})}\BibitemShut {NoStop}%
\bibitem [{\citenamefont {Tsirlin}\ \emph {et~al.}(2008)\citenamefont
  {Tsirlin}, \citenamefont {Belik}, \citenamefont {Shpanchenko}, \citenamefont
  {Antipov}, \citenamefont {Takayama-Muromachi},\ and\ \citenamefont
  {Rosner}}]{tsirlin2008}%
  \BibitemOpen
  \bibfield  {author} {\bibinfo {author} {\bibfnamefont {A.~A.}\ \bibnamefont
  {Tsirlin}}, \bibinfo {author} {\bibfnamefont {A.~A.}\ \bibnamefont {Belik}},
  \bibinfo {author} {\bibfnamefont {R.~V.}\ \bibnamefont {Shpanchenko}},
  \bibinfo {author} {\bibfnamefont {E.~V.}\ \bibnamefont {Antipov}}, \bibinfo
  {author} {\bibfnamefont {E.}~\bibnamefont {Takayama-Muromachi}}, \ and\
  \bibinfo {author} {\bibfnamefont {H.}~\bibnamefont {Rosner}},\ }\href@noop {}
  {\bibfield  {journal} {\bibinfo  {journal} {Phys. Rev. B}\ }\textbf {\bibinfo
  {volume} {77}},\ \bibinfo {pages} {092402} (\bibinfo {year}
  {2008})}\BibitemShut {NoStop}%
\bibitem [{\citenamefont {Oka}\ \emph {et~al.}(2008)\citenamefont {Oka},
  \citenamefont {Yamada}, \citenamefont {Azuma}, \citenamefont {Takeshita},
  \citenamefont {Satoh}, \citenamefont {Koda}, \citenamefont {Kadono},
  \citenamefont {Takano},\ and\ \citenamefont {Shimakawa}}]{oka2008}%
  \BibitemOpen
  \bibfield  {author} {\bibinfo {author} {\bibfnamefont {K.}~\bibnamefont
  {Oka}}, \bibinfo {author} {\bibfnamefont {I.}~\bibnamefont {Yamada}},
  \bibinfo {author} {\bibfnamefont {M.}~\bibnamefont {Azuma}}, \bibinfo
  {author} {\bibfnamefont {S.}~\bibnamefont {Takeshita}}, \bibinfo {author}
  {\bibfnamefont {K.~H.}\ \bibnamefont {Satoh}}, \bibinfo {author}
  {\bibfnamefont {A.}~\bibnamefont {Koda}}, \bibinfo {author} {\bibfnamefont
  {R.}~\bibnamefont {Kadono}}, \bibinfo {author} {\bibfnamefont
  {M.}~\bibnamefont {Takano}}, \ and\ \bibinfo {author} {\bibfnamefont
  {Y.}~\bibnamefont {Shimakawa}},\ }\href@noop {} {\bibfield  {journal}
  {\bibinfo  {journal} {Inorg. Chem.}\ }\textbf {\bibinfo {volume} {47}},\
  \bibinfo {pages} {7355} (\bibinfo {year} {2008})}\BibitemShut {NoStop}%
\end{thebibliography}
\end{document}